\documentclass{article}

\usepackage{amsmath,bm}
\usepackage{mathtools}

\usepackage{graphicx,subfigure}
\usepackage{multirow}
\usepackage{amsmath} % assumes amsmath package installed
\usepackage{amssymb}  % assumes amsmath package installed

\usepackage{xr}
\usepackage[superscript]{cite}

\graphicspath{{pic/}{figures/}}

\date{}
\title{Temporal scaling in information propagation}

\author{Junming Huang$^1$ \and Chao Li$^{1,2}$ \and Wen-Qiang Wang$^2$ \and Hua-Wei Shen$^1$ \and Guojie Li$^1$ \and Xue-Qi Cheng$^1$}

\begin{document}
\maketitle

%Junming Huang$^{1}$,
%Chao Li$^{1,2}$,
%Wen-Qiang Wang$^{2}$,
%Hua-Wei Shen$^{1,\ast}$,
%Xue-Qi Cheng$^{1}$,
%Yong-Yeol Ahn$^{2,\ast}$
%\\
\begin{flushleft}
1 Institute of Computing Technology, Chinese Academy of Sciences, Beijing, People's Republic of China
\\
2 Web Sciences Center, School of Computer Science and Engineering, University of Electronic Science and Technology of China, Chengdu, Sichuan, People's Republic of China
\\
$\ast$ Corresponding Author: shenhuawei@ict.ac.cn
\end{flushleft}

% intro
% (0) abs: looking for law; lacks dynamics; we find decay phenomenon; model proposed to improve prediction
\begin{abstract}
    For the study of information propagation, one fundamental problem is uncovering universal laws governing the dynamics of information propagation. This problem, from the microscopic perspective, is formulated as estimating the propagation probability that a piece of information propagates from one individual to another. Such a propagation probability generally depends on two major classes of factors: the intrinsic attractiveness of information and the interactions between individuals. Despite the fact that the temporal effect of attractiveness is widely studied, temporal laws underlying individual interactions remain unclear, causing inaccurate prediction of information propagation on evolving social networks.
    In this report, we empirically study the dynamics of information propagation, using the dataset from a population-scale social media website. We discover a temporal scaling in information propagation: the probability a message propagates between two individuals decays with the length of time latency since their latest interaction, obeying a power-law rule. Leveraging the scaling law, we further propose a temporal model to estimate future propagation probabilities between individuals, reducing the error rate of information propagation prediction from $6.7\%$ to $2.6\%$ and improving viral marketing with $9.7\%$ incremental customers.
\end{abstract}

In recent years, information propagation on social networks has been attracting much attention from academia and industry~\cite{background-im-2, individual-influence-1, background-influence-2, background-influence-3, background-influence-5, background-influence-6, background-influence-7, background-influence-8, background-influence-9}.
% understanding micro mechanism is important: understand intrinsics, guide applications such as IM [domingoes, kempe, jure, chen, cheng]
Understanding the mechanisms of information propagation, with or without exogenous and endogenous factors, is a fundamental task to uncover the universal laws governing the process of information propagation, which is important for better explaining the dynamics of information propagation~\cite{background-im-1}, predicting information popularity~\cite{empirical-bao}, and initiating viral marketing campaign~\cite{im-formalize,im-algorithm-CELF,im-algorithm-newgreedy,im-algorithm-CELF++,im-algorithm-staticgreedy}.
% formalized as propagation probability
This task, from the microscopic perspective, is formulated as inferring and estimating the propagation probability that a piece of information propagates from one individual to another along social links connecting them.

%(1.2) background: lacks research in dynamics of interactions
% propagation probability attribute to message and interaction.
The difficulty of estimating propagation probability lies in the complex interaction pattern between individuals and the co-existence of various confounding factors, such as the interplay between social selection and social influence. Previous studies empirically identified two classes of factors that drive information propagation: the attractiveness of information and the interactions between individuals.
%(1.2.1) message attractiveness: three factors.
	% time-invariant attractiveness [lerman], time-variance popularity (growing) [wang-barabasi], time-dependent freshness (decaying) [gomez, gomez, goyal, huberman, song, wang-barabasi]
Existing studies on the first class mainly discussed three fundamental mechanisms with respect to message attractiveness~\cite{message-attractiveness-wang}: the time-invariant intrinsic attractiveness or fitness~\cite{interestingness-lerman, content-ye}, the Matthew effect in the popularity accumulation~\cite{message-attractiveness-wang}, and the freshness of messages decaying in a power-law~\cite{behavior-gomez2010}, exponential~\cite{behavior-goyal,behavior-song}, Rayleigh~\cite{behavior-gomez2011,behavior-gomez2013}, or log-normal~\cite{message-attractiveness-wang} manner with respect to the time span since the message is posted~\cite{latency-1}.
%(1.2.2) individual interactions
	% typical works that analyze individual interaction to predict propagation probability include:
	% time-invariant frequency [saito, goyal] and  structure [lu, zhang, fang].
In contrast, most conventional studies on the second class were limited to static or quasi-static scenarios, assuming time-invariant interactions between any pair of individuals. Researchers estimated a propagation probability by indifferently aggregating recent and long-ago interactions~\cite{behavior-goyal, behavior-saito}, or by learning a probability function with static features including structural characteristics of the underlying network~\cite{empirical-bao, struture-zhang, struture-lu, content-fang}, demographic features~\cite{empirical-aral}, and topical and contextual features~\cite{empirical-suh, empirical-feng, empirical-yang}.
	% influence that might change with time. modeled in a Markovian chain. didn't reveal the temporal scaling. [wang]
	% time-dependent scaling: unknown
Few studies explored the possibility that individual interactions change with time. A recent study modeled social influence as a Markovian chain on temporally sliced snapshots of a social network, but did not reveal the intrinsic temporal scaling how social influence evolved~\cite{struture-wang}.

%	necessary to cover the hole: prob determined by instant interaction, neglecting the dynamics could be problematic; instant interaction can only be calculated with instant communications; but the most important task, predicting the future, has no communication data to use; therefore it is a must to explore the dynamics of interaction.
Actually, most real-world social networks are far from static. On evolving social networks, whether a piece of information will be propagated is more related to instant frequency of individual interactions rather than average frequency indifferently aggregated over recent and long-ago interactions. Hence, it is problematic to neglect the dynamic nature of individual interactions and its crucial role at information propagation, leading to inaccurate predictions. A possible solution is working only on recent interactions based on temporally sliced snapshots of interactions. However, it is hard to determine the appropriate temporal scale of snapshots since the frequency of interactions is scale-free~\cite{power-law-3}. Therefore, we lack a full understanding about the temporal scaling of information propagation, which is crucial to grasp the propagation dynamics of information.

%(1.3) specific problem: temporal scaling in individual interactions
% we address the problem: whether individual interaction varies, how to predict instant individual interaction in any future time?
In this report, we study whether and how individual interactions vary temporally and their role at predicting the instant propagation probability.
% intuitively, recent dense communication suggests strong interaction and thus high probability and vice versa (traditional belief)
% the delegate of recency is latency (idle time). with the latency grows, the recency window changes from dense to sparse, therefore instant probability decays.
Intuitively, a high frequency of recent communication implies strong instant interaction and a high propagation probability. As the delegate of recency, \emph{latency} is defined as the idle time since the latest communication between two individuals. A long latency generally reflects a low tendency of future interaction. Thus analyzing the interdependence between the latency and the trend of a propagation probability provides us a peculiar delegate for understanding the temporal effect of information propagation. With this delegate, we study on a population-scale social media dataset and conduct an empirical validation for the intuition that a longer latency indicates a relatively lower instant propagation probability.

% practically, we consider probability = interactions, because the message factor is averaged out
% in large-scale statistics, message-related factors are averaged out, therefore the temporal scaling of information equals to the temporal scaling of individual interactions
To focus on analyzing the temporal scaling of propagation probabilities from the perspective of individual interactions, in this report we do not consider the factors of information attractiveness, and instead calculate a propagation probability between two individuals as the ratio of retweeted and neglected messages that are propagated from one to another. This methodology is reasonable when the number of messages is sufficient to largely average out information attractiveness. In this way the temporal scaling of information propagation fully reflects the temporal scaling of individual interactions.

% results
\section*{Results}
% (2)results
% (2.0)data preparation and formulation
	% weibo (brief)
	% formalization
% (2.1) temporal scaling
	% (2.1.0) motivation: examine temporal scaling of information propagation
	% (2.1.1) empirical finding
	% time stamps on single edge -> non-uniform
	% time stamps on all edges -> burst
	% latency distribution: power-law
	% if the interactions are static, the latency should follow an exponential distribution.
	% (2.1.2) prob against latency
	% prob is temporal -> each behavior is associated with a unique instant prob
	% instant prob is related with instant individual interaction. we again use latency as delegate to investigate.
	% shorter latency leads to higher instant prob
	% estimate edge-free prob with ratio
	% empirical find prob decays with latency in a power law
	% to better fit heterogeneous edges, allows local decaying speed

%(2.1.0) data preparation and formulation
	% data weibo (brief)
The studies are based on a publicly available dataset (WISE 2012 Challenge, http://www.wise2012.cs.ucy.ac.cy/challenge.html) collected from Sina Weibo, the largest Chinese micro blogging website, like Twitter. In the dataset with some simple preprocessing (see Section S1), half a million users created $1.2$ million following relations among them, providing channels for propagation of $8$ million messages.
	% formalization	
	% latency definition: time span on a single edge
We denote with an edge $(v_i,v_j)$ the relation that a user $v_j$ (called the \emph{follower}) follows another user $v_i$ (called the \emph{followee}). Each time $v_j$ sees a message $k$ posted or retweeted by $v_i$ that $v_j$ has not retweeted before, we say $\delta_{i,j,k}=1$ if $v_j$ retweets $k$, forming a \emph{positive example} indicating $v_i$ successfully activates $v_j$ to retweet $k$; otherwise $\delta_{i,j,k}=0$ for a \emph{negative example} if $v_j$ neglects $k$. For each positive / negative example, we measure the \emph{latency} $\tau_{i,j,k}$ as the time span since the latest time $v_j$ retweets a message from $v_i$.

\begin{figure}[ht]
  \centering
  \includegraphics[width=\textwidth] {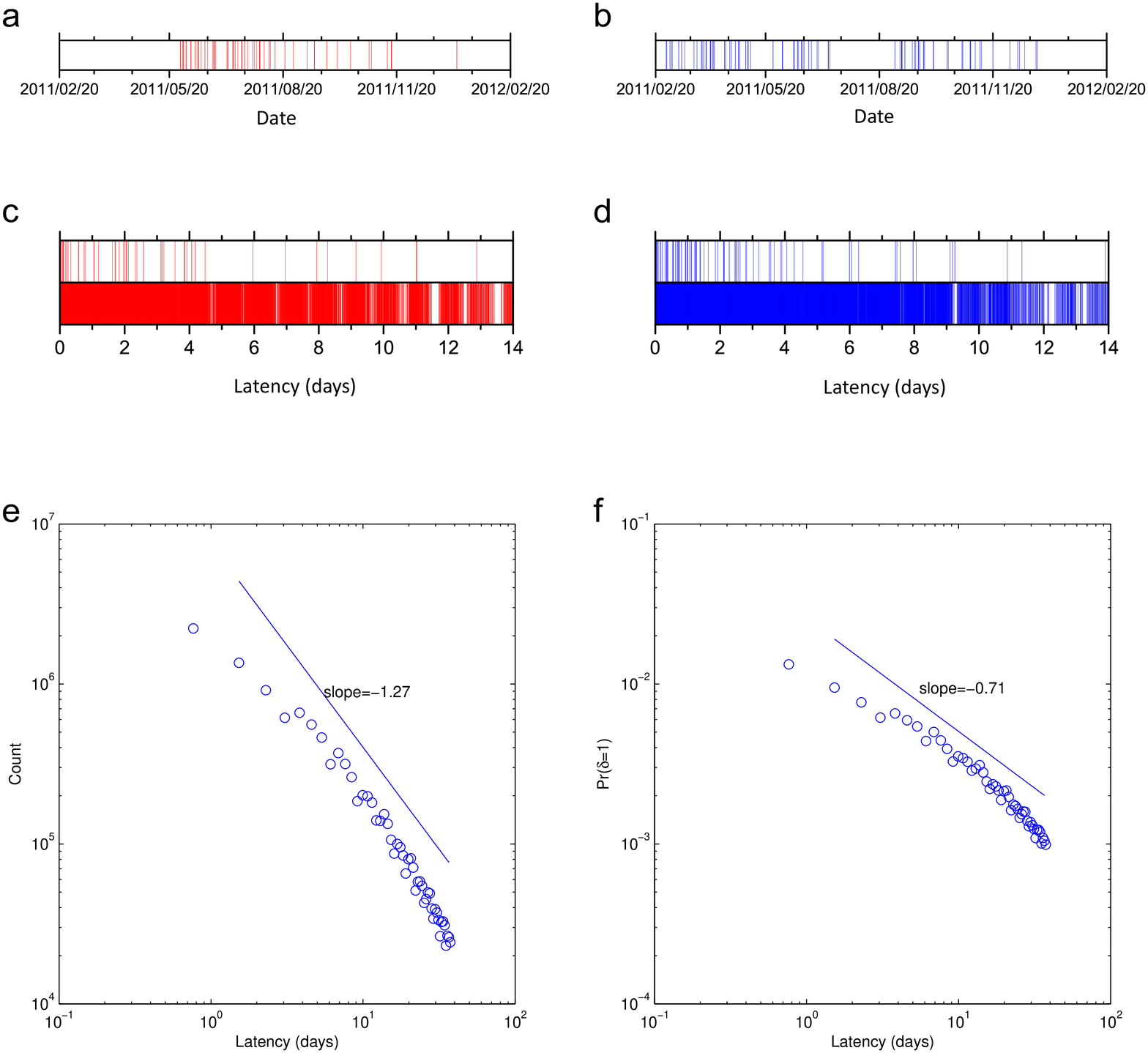}
%  \subfigure[]
%    {
%    \includegraphics[width=0.45 \textwidth]
%    {44477-1658690-succtime.eps}
%    \label{fig:typical.edge.A}
%    }
%  \subfigure[]
%    {
%    \includegraphics[width=0.45 \textwidth]
%    {44477-1155408-succtime.eps}
%    \label{fig:typical.edge.B}
%    }
%  \subfigure[]
%    {
%    \includegraphics[width=0.45 \textwidth]
%    {44477-1658690-intvl-14.eps}
%    \label{fig:pos.neg.examples.on.case.A}
%    }
%  \subfigure[]
%    {
%    \includegraphics[width=0.45 \textwidth]
%    {44477-1155408-intvl-14.eps}
%    \label{fig:pos.neg.examples.on.case.B}
%    }
%  \subfigure[]
%    {
%    \includegraphics[width=0.45 \textwidth]
%    {interval-distri-2013-09-16-13-12-54-positive-examples-first-half.eps}
%    \label{fig:interval.distri}
%    }
%  \subfigure[]
%    {
%    \includegraphics[width=0.45 \textwidth]
%    {interval-distri-2013-09-16-13-12-54-prob-first-half.eps}
%    \label{fig:prob.against.interval}
%    }
  \caption{\textbf{Characterizing propagation probabilities.} \textbf{(a,b)} Time stamps of positive examples (retweeting behaviors) on two random edges. Each vertical line represents a retweeting behaviors occurring with the time stamp marked on the horizontal axis. \textbf{(c,d)} Positive (retweeting) and negative (neglecting) examples on those two edges. Vertical lines in upper half represent positive examples, while those in lower half represent negative ones. It shows an obvious tendency that most positive examples are concentrated on the left zone, i.e., most retweeting behaviors occur with short latency. The tendency is stronger on (c) than that on (d). \textbf{(e)} Distribution of latency of retweeting behaviors over all edges. \textbf{(f)} Ratio of positive examples upon all examples on all edges with respect to the associated latency, demonstrating the power-law interdependence between the propagation probability and the latency.}
  \label{fig:case}
\end{figure}

% (2.1) temporal scaling
	% (2.1.0) motivation: examine temporal scaling of information propagation
	% (2.1.1) empirical finding
	% time stamps on single edge -> non-uniform
	% time stamps on all edges -> burst
	% latency distribution: power-law
	% if the interactions are static, the latency should follow an exponential distribution.
We start to explore the temporal scaling of information propagation by examining time stamps of positive examples on two randomly selected edges, a followee and two of his followers. %, $edge (v_i=44477,v_j=1658690)$ and $edge (v_i=44477,v_{j'}=1155408)$.
Figure~\ref{fig:case}a and Figure~\ref{fig:case}b reveal a non-uniform density of positive examples that the followers frequently retweet messages from the followee in several short time periods, separated by long idle periods. This implies a \emph{burst} phenomenon on individual interactions: short time frames of intense interactions are separated by long idle periods~\cite{power-law-3}. To provide a solid evidence for the existence of burst in retweeting behaviors, we depict in Figure~\ref{fig:case}e the distribution of latency of all positive examples. The power-law distribution of latency, reflecting the emergence of bursty retweeting behaviors, exhibits the temporal nature of individual interactions. Note that static individual interactions lead to a time-invariant propagation probability on each edge in this scenario, which views retweeting behaviors as a homogeneous Poisson process, resulting in an exponential distribution of latency.

	% (2.1.2) prob against latency
	% prob is temporal -> each behavior is associated with a unique instant prob
	% instant prob is related with instant individual interaction. we again use latency as delegate to investigate.
	% shorter latency leads to higher instant prob
	% estimate edge-free prob with ratio
	% empirical find prob decays with latency in a power law
	% to better fit heterogeneous edges, allows local decaying speed
The temporal nature of individual interactions results in a necessity to assign a unique propagation probability to every retweeting / neglecting behavior even occurred on the same edge, reflecting the instant tendency that a follower retweets a followee's message at the time that message arrives. To uncover the temporal scaling of instant propagation probabilities, we investigate the interdependence between the propagation probability behind every retweeting / neglecting behavior and the latency associated with it. The interdependence is suggested by the distribution of retweeting / neglecting behaviors on those two edges against associated latency, where most retweeting behaviors occur with short latency (Figure~\ref{fig:case}c and~\ref{fig:case}d). We calculate the ratio of retweeting and neglecting behaviors over all edges to estimate the invisible instant propagation probability given certain latency. The propagation probability decreases with the latency in a power-law manner (Figure~\ref{fig:case}f).  % , supporting the hypothesis that propagation probabilities changes with time intervals and simultaneously refusing the alternative explanation of static propagation probabilities.
Fitting the log-log curve in Figure~\ref{fig:case}f produces a consistently decaying speed of $-0.71$ slope, suggesting the temporal scaling between a propagation probability $Pr(\delta=1)$ behind a retweeting / neglecting behavior and its associated latency $\tau$ as follows,
\begin{equation}\label{equ-global-decay}
Pr(\delta_{i,j,k}=1) \propto \tau_{i,j,k}^{-0.71}.
\end{equation}

We further study whether retweeting behaviors on different edges share the same power exponent, governing the temporal scaling. As shown in Figure~\ref{fig:case}a-d, although the retweeting behaviors on the two edges both obey the power-law temporal scaling, the power exponents are quite different. Therefore, we need to assign an edge-specific exponent on each edge, in order to model the temporal scaling of information propagation on various edges of social networks.

%\begin{figure}[ht]
%  \centering
%    \includegraphics[width=0.45 \textwidth]
%    {Huang-1.eps}
%  \caption{Empirical study. Figure~\ref{fig:case}a and Figure~\ref{fig:case}b show retweet time stamps on two typical edges. Each vertical line represents a retweeting behavior whose time stamp is marked in the horizontal axis. Figure~\ref{fig:case}e demonstrate the distribution of intervals over all edges. Figure~\ref{fig:case}c and Figure~\ref{fig:case}d show positive (retweeting) and negative (neglecting) examples on two typical edges. Vertical lines in upper half represent positive examples, while vertical lines in lower half represent negative ones. It shows an obvious tendency that most positive examples are concentrated on the left, i.e., most retweeting behaviors occur in short intervals. The tendency is even more significant on Figure~\ref{fig:case}c. Figure~\ref{fig:case}f demonstrates how the propagation probability changes against the interval.}
%\end{figure}

%\subsection{Prediction performance}

Motivated by the observed temporal scaling, we propose a temporal model, namely \emph{Decay model}, to predict propagation probability. We evaluate the performance of the model by applying it to predict retweeting behaviors and to launch a viral marketing strategy, compared with four mainstream baselines, namely MLE, EM~\cite{behavior-saito}, Static Bernoulli~\cite{behavior-goyal}, and Static PC Bernoulli~\cite{behavior-goyal}.

\begin{figure}[ht]
\centering
\includegraphics[width= \textwidth]{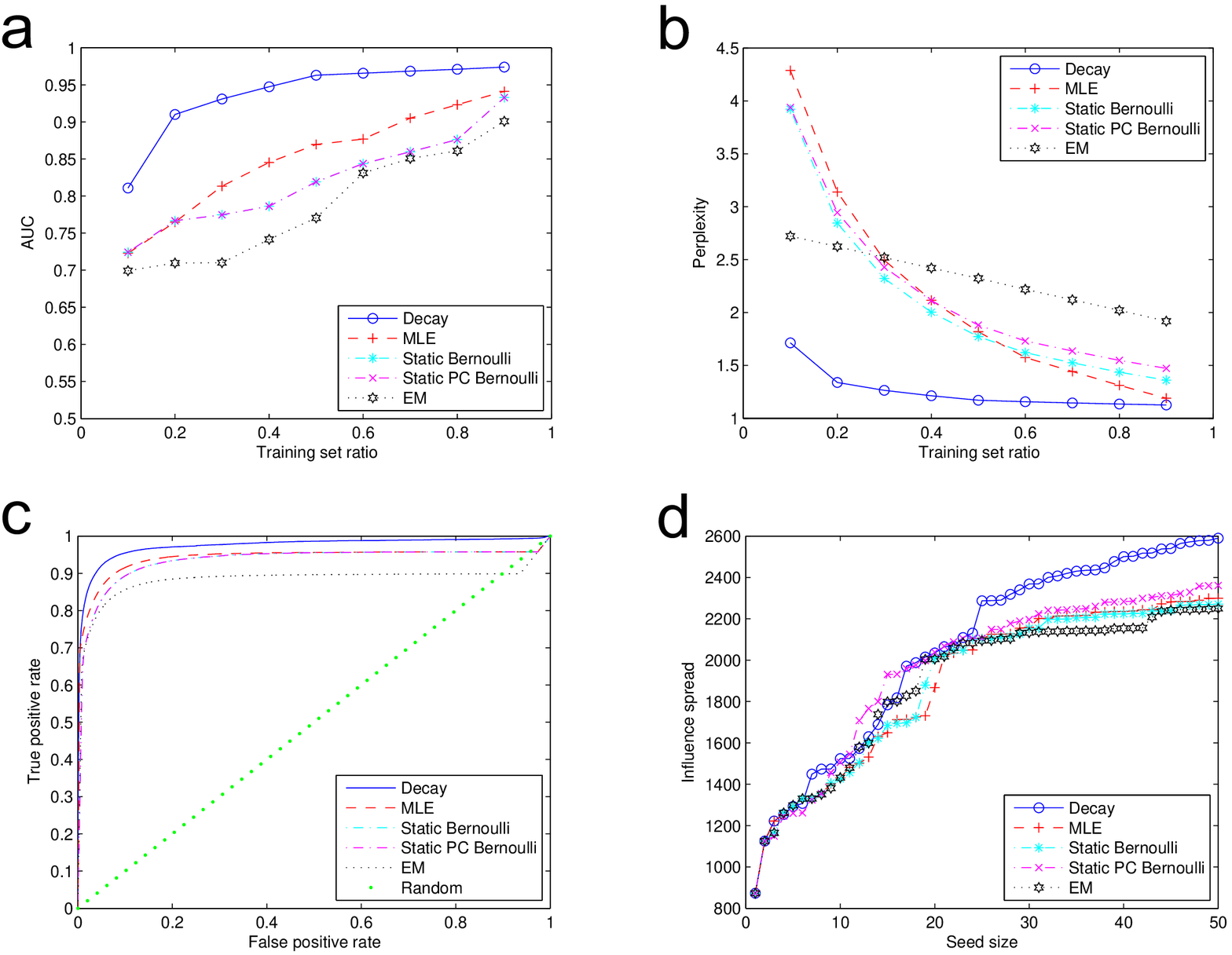}
\caption{
%(Color online)
\textbf{Model evaluation}. % and designing viral marketing strategies.
\textbf{(a)} AUC of the Decay model and baselines. AUC measures the area under the ROC curves, and thus is equivalent to the probability that a trained model correctly distinguish a randomly selected positive example from another randomly selected negative example. \textbf{(b)} Perplexity of the Decay model and baselines when predicting retweeting behaviors, against the training set ratio. A lower perplexity indicates a better prediction accuracy, meaning less extent a testing example surprises a trained model. \textbf{(c)} Receiver Operating Characteristic (ROC) curves with a training set of $90\%$ examples.  \textbf{(d)} Influence spreads of an initial seed set selected on propagation probabilities predicted by the Decay model and baselines.
}
\label{fig:eval}
\end{figure}

% \subsubsection{Retweeting prediction}
% \textbf{Retweeting prediction} ...
The first evaluation experiment measures the probability a model correctly predicts whether or not an individual will retweet an incoming message. Figure~\ref{fig:eval}a reports AUC, the area under the Receiver Operating Characteristic (ROC) curve, equivalent to the probability that a classifier correctly distinguishes a positive example from a negative one. The Decay model outperforms all baselines, raising AUC from $93.3\%$ to more than $97.4\%$. Intuitively speaking, when facing a randomly selected pair of a retweeting behavior and a neglecting behavior, the error rate to incorrectly distinguish them is reduced by a half by the Decay model over the best baseline. We then report the \emph{perplexity} on the testing set against the training set ratio to obtain the probability that a model, trained with incomplete observations, correctly generates the testing examples. As shown in Figure~\ref{fig:eval}b, the Decay model achieves the lowest (best) perplexity among all tested models. The priority of the Decay model is consistent in all examined training set ratios, with a more significant improvement on a relatively smaller training set. We also evaluate the Decay model with ROC curve, which is a metric appropriate for extremely imbalanced datasets such as the one we use in this report (as well as most real-world social media) where positive examples occupy less than $1\%$. ROC, measuring the sensitivity (true positive rate) against specificity (one minus false positive rate), is insensitive to the ratio between positive and negative examples. Figure~\ref{fig:eval}c reports the ROC curves of the Decay model and baselines with $90\%$ examples held out as the training set. Results of other training set ratios are similar. The figure shows that the Decay model achieves the best capability at distinguishing retweeting behaviors from neglecting behaviors with a significant improvement upon all baselines.

The second evaluation measures the accuracy a model predicts propagation probabilities. Intuitively, predictions that are more accurate would help select a better initial seed set, triggering a larger fraction of individuals.
% \textbf{Viral marketing} is an important application of information propagation, which is usually formalized as the influence maximization problem that aims to trigger a maximized spread from a carefully selected seed set based on given propagation probabilities on all edges of a social network. Various methods have already been proposed to efficiently find a near-optimal set of initial seed nodes to maximize the final spread, which is defined as the number of nodes eventually activated from seed nodes. Therefore the influence maximization problem provides a good application scenario to evaluate propagation probability estimation methods, since accurate estimations better predict the propagation and therefore help select a better seed set, indicated with a larger spread.
We split all examples into $4$ groups in a chronological order with respect to example time stamps. Each group contains examples in $30$ weeks (see Section S6 for details). The Decay model and baselines train on examples in the earlier $205$ days (training phase) and predict the propagation probabilities in the last $5$ days (evaluation phase). Based on those predictions, a state-of-the-art influence maximization algorithm (CELF++~\cite{im-algorithm-CELF++}) is used to select an initial seed set maximizing the expected eventual influence spread. We then estimate the \emph{pseudo actual} spread of such a seed set as the number of nodes reachable from the seed set on a propagation network, which is a subgraph of the social network consisting of edges with at least one actual retweeting behavior in the last $5$ days. As reported in Figure~\ref{fig:eval}d (one group shown only), the largest pseudo actual spread comes from the seed set selected on propagation probabilities predicted by the Decay model, which eventually reaches $2,590$ nodes, achieving a $9.7\%$ increase upon what is reached by the best baseline, i.e., Static PC Bernoulli which reaches $2,361$ nodes.
The increase in pseudo actual spread demonstrates the advantage that the Decay model more accurately predicts the propagation probabilities, confirming our finding that individual interactions decay with latency.

% discussion
\section*{Discussion}
%(3) discussions
% main conclusion: find a rule; build a prediction model
%machine learning perspective
%q as a free variable or function
%multiple testing examples in sequence
%open topics: dynamic network, mechanism
%we beat the explanation that "the power-law distribution of retweet interval is a trivial result of the power-law distribution of post interval", which would result in a constant probability.

%(3) discussions

% main conclusion: find a rule; build a prediction model
In this report, we uncovered the temporal scaling in information propagation from the perspective of individual interactions: a propagation probability decays slowly in a power-law manner with the latency since their latest interaction. Such a dynamic nature was demonstrated by empirical studies on a large-scale public social media dataset, showing the power-law interdependence between a propagation probability and latency.

% ignoring the dynamic is unwise
With the observed temporal scaling, a Decay model was proposed to predict future propagation probability among individuals, incorporating a time-invariant base probability and a time-decaying exponent on each edge. The model is applicable in scenarios where an underlying social network and tractable information propagation with time stamps are observed, such as micro blogging (Twitter and Sina Weibo), blog sites, book sharing sites and email promotion networks. Empirical evaluations supported that the Decay model outperformed mainstream baselines in predicting retweeting behaviors, significantly reducing by a half the expected error rate of incorrectly identifying a retweeting behavior.

%machine learning perspective. no suffering sparsity.
From the perspective of machine learning, the discovered temporal scaling provides an additional feature to estimate propagation probability. While traditional models assume static propagation probability, the proposed Decay model additionally explores the temporal effect of a propagation probability, explaining the increased accuracy. %While traditional models consider only one predictor that the ratio of positive and negative examples, the proposed temporal model considers two predictors, the base probability and the interval. The increased number of predictors leveraged leads increased accuracy.
Generally speaking, a model with more features requiring more data for training suffers severe over-fitting problem on sparse data. This partly explains why traditional models do not consider temporal features. In order to reduce the pain of sparsity, the Decay model introduces a prior distribution of the decaying exponent $p(\alpha)$, suggested by the global decaying exponent in empirical study results. The prior distribution successfully reduces the pain of sparsity: the improvement of the Decay model upon baselines is even more significant with a relatively smaller training set (Figure~\ref{fig:eval}a and~\ref{fig:eval}b). Note that typically only several retweeting behaviors are observed on an edge in a real-world scenario, the outstanding performance of the Decay model on sparse data is of great importance in practice.

% IC
%Between two well-known models that explain information propagation process, namely Independent Cascade model and Linear Threshold model, in this report we discuss temporal mechanism of information propagation with propagation probabilities under the former model. With the typical setting of the Independent Cascade model that each individual can not be activated multiple times with the same message, we consider a follower $v_j$ decides whether to retweet a message $k$ posted or retweeted by his followee $v_i$ only when $v_j$ has not retweeted it before. In fact, most discussions applicable to the Independent Cascade model are also applicable to the Linear Threshold model, as Kempe et al. suggested~\cite{im-formalize}.

% IM
It is worth noting that the viral marketing evaluation is not conducted using Monte Carlo simulations, as done in most influence maximization studies. That is because what we compare is the configurations of propagation probabilities estimated with various model, and thus it is unfair to run Monte Carlo simulations with any estimated configuration, otherwise estimating all probabilities equal to one will surely win. Instead, we estimate the propagation spread in a pseudo-actual way. We build a propagation network, a subgraph of the social network, with edges where at least one retweeting behavior occurs in the $5$-day evaluation phase. Therefore the reachability of a node on the propagation network measures its pseudo actual influence spread during that $5$ days. It is equivalent to one Monte Carlo simulation that is produced from the (unknown) actual individuals and observed by actual retweeting behaviors. The estimated propagation spread is deterministic without any random deviation.

%q as a free variable or function
In the Decay model, the base probability $q$ is considered as a free variable whose value is fully determined by maximum-a-posteriori inference with a prior distribution. In fact, the Decay model can certainly incorporate any endogenous or exogenous factors through rewriting $q$ as a function of those factors, such as demographical, structural, content and context features. Parameters of such a function could also be estimated in maximum-a-posteriori inference.

%multiple testing examples in sequence
In the first evaluation experiment, the Decay model is tested with only one testing example on each edge, for the ease of calculating latency. When facing multiple testing examples (e.g., predicting whether an individual will retweet a series of messages in a month), one should predict those examples one by one in a chronological order and calculate the expected latency of a later example over the joint probability distribution of predicted results of all previous testing examples.

% Markovian
Choosing the latency as a delegate of recency is equivalent to approximating the information propagation occurrences as a first order Markov process, i.e., only the idle time since the latest interaction, instead of all historical interactions, affects the current decision. Such an approximation, effectively avoiding expensive calculation with an nondeterministic number of parameters required to build a complicated function defined on all historical interactions, succeeds in revealing strong evidence of interdependence between propagation probabilities and latency and in building an outperforming prediction model. That supports the important role that the temporal scaling plays in characterizing a propagation probability.

%we beat the explanation that "the power-law distribution of retweet interval is a trivial result of the power-law distribution of post interval", which would result in a constant probability.

% IM: That also reveals that in real-world scenarios propagation probabilities vary with time, therefore a model assuming static probabilities loses prediction accuracy.

%open topics: evolving influential nodes
As an open question in future, it would be attractive to characterizing influential nodes identified with high propagation probabilities estimated by the Decay model, and to demonstrate the evolving distribution of instant influential nodes on a social network.

\section*{Methods}
%\subsection{Decay model}

The proposed \emph{Decay model} describes the propagation probability $P(\delta_{i,j,k}=1)$, that an individual $v_i$ will successfully activate another individual $v_j$ to retweet a message $k$, which is believed to be determined by two factors:

\begin{itemize}
    \item $q_{i,j} \in [0,1]$: the base probability associated with the edge $(v_i,v_j)$;
    \item $\tau_{i,j,k} \in [1,+\infty)$: latency, the time span since the latest time $v_i$ activated $v_j$, i.e., $\tau_{i,j,k}=t_{k,i}-t_{k',j}$, where $t_{k,i}$ is the time stamp when $v_i$ posts or retweets $k$, and $k'$ is the latest message before $k$ that $v_i$ activates $v_j$ to retweet.
\end{itemize}

Specifically, the propagation probability is as follows,
\begin{equation}\label{equ:decay-model}
P(\delta_{i,j,k}=1) = q_{i,j} \tau_{i,j,k}^{-\alpha_{i,j}}, %, \tau_{i,j,k} \geq 1,
\end{equation}
where $\alpha_{i,j} > 0$ is a decaying exponent associated with the edge $(v_i,v_j)$. The decaying exponent is edge-specific, with a prior distribution $p(\alpha)$ reflecting the global decaying exponent. Traditional models without temporal scaling of propagation probabilities can be viewed as special cases of the Decay model with constant $\alpha_{\cdot,\cdot} = 0$.

Latency is required to be bounded, i.e., $\tau \geq 1$, to guarantee $P(\delta_{i,j,k}=1) \in [0,1]$. Specifically, $\tau_{i,j,k}=1$ results in that $q_{i,j} \tau_{i,j,k}^{-\alpha_{i,j}} = q_{i,j}$, revealing the intuitive meaning of the base probability that $q_{i,j}$ equals to the probability $v_i$ successfully activate $v_j$ to retweet a message $k$ which arrives immediately after a previous successful activation.

%\subsection{inference}
The hidden parameters $q$ and $\alpha$ are inferred with a maximum-a-posteriori estimate with prior distributions $p(q)$ and $p(\alpha)$. See Section S3 for details.

% baselines
To demonstrate the performance of the Decay model, four mainstream baselines are implemented to estimate and predict propagation probabilities on all edges, including MLE, EM~\cite{behavior-saito}, Static Bernoulli~\cite{behavior-goyal}, and Static PC Bernoulli~\cite{behavior-goyal} (see Section~S4). Some other widely used models are not compared because those models require user profiles or message content that are absent in this scenario.

% experiment setup: retweeting prediction
In the retweeting prediction experiment, we apply a next-one strategy to split a training set and a testing set. On each edge, we sort all examples in a chronological order, take the earliest $N\%$ examples as the training set, and leave the next one example as the testing set. Thus the size of the training set increases with $N\%$, the \emph{training set ratio}, while the size of the testing set is a constant equal to the number of edges. With parameters trained on the training set, the Decay model predicts the label $\delta$ of examples in the testing set.

% evaluation metrics
The evaluation metrics include \emph{perplexity}, \emph{ROC} curve and \emph{AUC}. The perplexity measures how the testing examples surprise a trained model. A lower perplexity demonstrates better prediction ability.
\begin{equation}
perplexity = e^{ -\frac
    {\sum_{\{v_i,v_j,k\} \in D_{test}} \delta_{i,j,k} \ln \tilde{P}(\delta_{i,j,k}=1) + (1 - \delta_{i,j,k}) \ln (1-\tilde{P}(\delta_{i,j,k}=1))}
    {|D_{test}|}
    }.
\end{equation}
where $D_{test}$ represents the testing set, and $\tilde{P}(\delta_{i,j,k}=1)$ is the estimated propagation probability. The Receiver Operating Characteristic (ROC) curve plots sensitivity (true positive rate) against specificity (one minus false positive rate). AUC measures the area under the Receiver Operating Characteristic curve, which is equivalent to the probability that a model correctly distinguishes a randomly selected positive example from a randomly selected negative example. A higher AUC indicates a better distinguish ability. See Section S5 for details.

\footnotesize{
\section*{Acknowledgments}
We thank Tao Zhou, Yanyan Lan, Suqi Cheng and Ming Tang for valuable discussions. This work was funded by the National Basic Research Program of China (973 Program) under grant number 2014CB340401, the National High-tech R\&D Program of China (863 Program) under grant number 2014AA015103, and the National Natural Science Foundation of China under grant numbers 61232010, 61202215, 61272536.
}

\footnotesize{
\section*{Author Contribution Statement}
JH and HWS designed research. JH, CL and WQW performed experiments. JH, CL, WQW, HWS, GL and XQC wrote and reviewed the manuscript.
}

\footnotesize{
\section*{Additional Information}
Competing financial interests: The authors declare no competing financial interests.
}
% References
\footnotesize{
%\bibliographystyle{naturemag}
%\bibliography{influ-infer}  % the name of the Bibliography .bib file

}
%\bibliographystyle{abbrv}
%\bibliographystyle{unsrt}
%\bibliography{influ-infer}  % the name of the Bibliography .bib file

%\balancecolumns
% That's all folks!
\end{document}